\def\Rpersquare{R_{\Box}}
\def\b{\bibitem}
\documentstyle[aps,prb,psfig,floats]{revtex} 
\begin{document}
\def\SNG{{\em Physical Review Style and Notation Guide}}
\def\LUG {{\em \LaTeX{} User's Guide \& Reference Manual}}
\def\btt#1{{\tt$\backslash$\string#1}}%
\def\REVTeX{REV\TeX}
\def\AmS{{\protect\the\textfont2
        A\kern-.1667em\lower.5ex\hbox{M}\kern-.125emS}}
\def\AmSLaTeX{\AmS-\LaTeX}
\def\BibTeX{\rm B{\sc ib}\TeX}
\twocolumn[\hsize\textwidth\columnwidth\hsize\csname@twocolumnfalse%
\endcsname
\title{Possible triplet superconductivity in MOSFETs}
\author{D. Belitz}
\address{Department of Physics and Materials Science Institute,
         University of Oregon,
         Eugene, OR 97403}
\author{T.R. Kirkpatrick}
\address{Institute for Physical Science and Technology, and Department of 
                                                                   Physics\\
         University of Maryland, College Park, MD 20742}
\date{\today}
\maketitle
\begin{abstract}
A theory that predicts a spin-triplet, even-parity superconducting ground
state in two-dimensional electron systems is re-analyzed in the light of
recent experiments showing a possible insulator-to-conductor transition
in such systems.
It is shown that the observations are consistent with such an exotic
superconductivity mechanism, and predictions are made for experiments that
would further corroborate or refute this proposal.
\end{abstract}
\pacs{PACS numbers: 74.20.-z , 64.60.Ak} 
]
Since 1979 it had been generally believed that two-dimensional ($2$-$d$)
electron systems cannot undergo a true metal-insulator transition (MIT), but
rather are always insulating at zero temperature ($T=0$), even for
arbitrarily weak disorder. Inititially, this was the conclusion only for
models of noninteracting electrons,\cite{g4} but later 
developments\cite{aal,f,R}
strongly suggested that it remains valid in the presence of electron-electron
interactions. Recent experiments on Si
MOSFETs have challenged this conventional wisdom.\cite{k,kPRL,kRC,pfw}
The samples used in
these experiments are in an unprecedented parameter regime, as they achieve
high electron mobilities at low electron densities. The apparent quantum
phase transition from an insulating phase to a conducting one that is
observed in these samples occurs at very low densities, which leads to a
very strong effective electron-electron interaction. The disorder is also
very strong, as is indicated by a `separatrix' value of the sheet resistance
that is about three times the Mott number. A MIT has also been observed
in p-type SiGe quantum wells with parameter values that are very close to
those in the MOSFETs.\cite{ottawa}

These observations can in principle be theoretically explained in a number 
of ways. For example, it is known that there is a MIT in $2$-$d$
systems with spin-orbit scattering and either no interaction, or a 
short-ranged interaction between the electrons.\cite{R}
While this universality class has been invoked in
the context of the experiments in question,\cite{p}
it is unclear why it should apply to the systems under consideration,
where the electrons interact by means of a strong, long-ranged Coulomb
interaction. Other possible theoretical explanations include
superconductor-to-insulator
transitions (SITs), which are well known to occur in $d=2$.\cite{R} Indeed, 
it has been
noted in Refs.\ \onlinecite{kPRL} and \onlinecite{kRC} that the
observed transition has many features in common with SITs. It has also
been pointed out in Refs.\ \onlinecite{pp-zr} and \onlinecite{vd} that
the conducting phase is likely to be a superconductor, or a perfect
conductor, respectively.

Some time ago the present authors have discussed a mechanism that can
lead to spin-triplet, even-parity superconductivity in disordered,
interacting electron systems.\cite{triplet}
The predicted effect is strongest in $d=2$,
and MOSFETs were explicitly mentioned in Ref.\ \onlinecite{triplet} 
as potential realizations of this phenomenon. This motivates us to
analyze this mechanism semi-quantitatively for parameter values that
are appropriate for the new MOSFET samples. As we will
show below, the observations are compatible with an explanation in
terms of superconducting fluctuations that preceed the predicted exotic 
superconducting state, and  we are also 
able to make predictions for observables other than the resistivity.

The physical mechanism that underlies the exotic
superconductivity discussed in Ref.\ \onlinecite{triplet} is based on
the slow decay of charge and spin polarization clouds that is caused
by long-time tail effects in disordered systems. The details have been
explained in Ref.\ \onlinecite{triplet}, and need not be repeated here.
We just recall that there are two physically distinct mechanisms, based
on charge and spin fluctuations, respectively.
Characteristic features of the ensuing even-parity, spin-triplet
superconductivity are,
(1) a very low $T_c$ for parameter values that
are characteristic of conventional $2$-$d$ electron systems, (2) a very
rapid variation of $T_c$ as a function of parameter values such as
the bare interaction amplitudes and the disorder, (3) a gapless
tunneling density of states, and (4) a logarithmic dependence of the
specific heat coefficient on the temperature. All of these features were
derived and discussed in Ref.\ \onlinecite{triplet}. In addition, one finds
(5) an ordinary Meissner effect, like in a gapless 
BCS superconductor, as we will now 
proceed to show.  To this end, we calculate the transverse
current-current susceptibility, using the field-theoretic formulation
of Ref.\ \onlinecite{triplet}. A standard calculation at the same level
as our previous saddle-point theory for the order parameter yields a
transverse current susceptibility
\begin{mathletters}
\label{eqs:2.2}
\begin{eqnarray}
\chi_T({\bf k}\rightarrow 0) = \frac{-2}{dm^2}\,\frac{1}{V}\sum_{\bf p} 
   {\bf p}^2\,T\sum_{n} \left[{\cal G}_n({\bf p})\,{\cal G}_n({\bf p})\right.
\nonumber\\
   - \left.{\cal F}_n({\bf p})\,{\cal F}_{-n}({\bf p})\right]\quad.
\label{eq:2.2a}
\end{eqnarray}
Here ${\cal G}$ and ${\cal F}$ are the normal and anomalous Green functions,
respectively,
\begin{eqnarray}
{\cal G}_n({\bf p}) = \frac{i\omega_n + \xi_{\bf p}}{\omega_n^2 + \xi_{\bf p}^2
                     - \Delta_n\Delta_{-n}}\quad,
\nonumber\\
{\cal F}_n({\bf p}) = \frac{\Delta_n}{\omega_n^2 + \xi_{\bf p}^2
                     - \Delta_n\Delta_{-n}}\quad,
\label{eq:2.2b}
\end{eqnarray}
with $\omega_n$ a fermionic Matsubara frequency, $\xi_{\bf p} = {\bf p}^2/2m
- \epsilon_F$ with $\epsilon_F$ the Fermi energy, and $\Delta_n$ the
frequency dependent gap function. The gap function on the imaginary axis
is purely real, $\Delta_n = -\Delta_{-n} = \Delta (i\omega_n) \equiv 
\delta (\omega_n)$,
with $\delta (\omega)$ a real function.\cite{reality_footnote}
At zero temperature, Eq.\ (\ref{eq:2.2a})
can then be expressed as a frequency integral,
\begin{equation}
\chi_T({\bf k}\rightarrow 0) = \frac{n}{m}\,\left[1 - \int_{0}^{\infty} d\omega
   \,\frac{\delta^2(\omega)}{\left(\omega^2 + \delta^2(\omega)\right)^{3/2}}
       \right]\quad,
\label{eq:2.2c}
\end{equation}
\end{mathletters}%
with $n$ the electron number density. We see that the correction to the
normal-metal result, $n/m$, is negative definite.\cite{asymptotics_footnote}
Consequently, we have 
$\chi_T({\bf k}\rightarrow 0) < n/m$, which guarantees ideal diamagnetism.

Si MOSFETs were discussed in Ref.\ \onlinecite{triplet} as possible
realizations of the proposed mechanism, but for typical parameter
values at best ultralow values for $T_c$, in the $\mu$K region, were to
be expected. For the samples investigated in 
Refs.\ \onlinecite{k,kPRL,kRC,pfw}, the situation is very
different. Let us
estimate the mean-field $T_c$ for these samples, assuming the charge
fluctuation mechanism described in Ref.\ \onlinecite{triplet} (we will
come back to the spin fluctuation mechanism later).
The mean-field value of $T_c$ is given by
\begin{equation}
T_c = T_F\,\frac{({\bar\kappa}/k_F)^2}{\hat\Rpersquare}\ 
          \exp\left(\frac{-4\pi^2/\ln 2}{\hat\Rpersquare}\right)\quad.
\label{eq:3.1}
\end{equation}
Here $\bar\kappa = \kappa (1 + F_0^s)$, with $\kappa$ the Thomas-Fermi 
screening wavenumber, and $F_0^s$ a Landau parameter. $T_F$ is the Fermi
temperature, and ${\hat\Rpersquare}$ is the resistance per square measured 
in units of $\hbar/e^2 = 4108\Omega$. With an electron density
$n = 10^{11}\ {\rm cm}^{-2}$ and a valley degeneracy $g_V = 2$, we have a Fermi
wavenumber $k_F = 5.6\times 10^5\ {\rm cm}^{-1}$. With an effective mass
$m^* = 0.2\ m_e$, this corresponds to a Fermi temperature $T_F = 6.9\ {\rm K}$.
Assuming a Landau parameter $F_0^s = -0.9$, the screening wavenumber is
$\bar\kappa = 1.5\times 10^7\ {\rm cm}^{-1}$. At a sheet resistance equal to
$R = 0.8\times 2\pi\hbar/e^2$, the $T_c$ formula, Eq.\ (\ref{eq:3.1}),
yields $T_c \approx 1.7\times 10^{-3}T_F \approx 10\ {\rm mK}$.

In the experimental range of Refs. \onlinecite{k,kPRL,kRC,pfw}, 
one therefore does not expect to see
true superconductivity resulting from the charge fluctuation mechanism.
If the observed transport anomalies are to be
explained in terms of it, then they must be caused
by superconducting fluctuations. Let us therefore estimate the width of
the temperature region within which one expects substantial fluctuations.
To this end we recall that the kernel in the gap or $T_c$ equation in
Ref.\ \onlinecite{triplet} takes the form of a function $F$,
whose argument depends on $T$ and ${\hat\Rpersquare}$. For 
asymptotically small arguments, $F$ approaches a logarithm, and we have
\begin{equation}
F\bigl({\hat\Rpersquare}(k_F/{\bar\kappa})^2 T/T_F\bigr) \approx 
    \ln \bigl({\hat\Rpersquare}(k_F/{\bar\kappa})^2 T/T_F\bigr) \equiv f(T).
\label{eq:3.2}
\end{equation}
If we consider $T$ the flow parameter in a RG description
of the system, then ${\hat\Rpersquare}$ becomes $T$-dependent, 
and for fixed bare
parameters $F$ can be considered a function $f$ of $T$, as is indicated
in Eq.\ (\ref{eq:3.2}).
If we replace $T$ by $T_c$, and use the above parameter values, then
we obtain $f(T_c)\approx -10$. As a criterion for noticeable fluctuations,
we require the temperature to be lower than some $T^*$ defined by 
$f(T^*) = -5$.
As the temperature increases from $T_c$, the scale dependent quantity
${\hat\Rpersquare}$ increases, but in $d=2$ it does so only 
logarithmically.\cite{R} The most important $T$-dependence is therefore
the explicit one in Eq.\ (\ref{eq:3.2}). Keeping only 
this leading $T$-dependence, the estimate becomes structurally the same
as in BCS theory.
Using the same parameters as for our estimate of $T_c$ above, we find
\begin{equation}
T^*/T_c \approx e^5 \approx 10^2\quad.
\label{eq:3.3}
\end{equation}
We also have obtained an estimate
of the same order by using the RG flow equation for the disorder 
parameter,\cite{R} which determines the $T$-dependence of the resistivity.

With the above values for $T_c$ and $T^*$, in a $2$-$d$ system with parameter
values appropriate for the new MOSFETs, Eq.\ (\ref{eq:3.3}) predicts that
observable superconducting fluctuations are to be expected up to
temperatures of about 1K, and the fluctuations should be substantial at the
lowest temperatures reached in the experiments. This could well explain the
observations. We stress that these considerations amount at best
to an order-of-magnitude estimate, while reliable calculations of
the superconducting $T_c$ are notoriously difficult for any mechanism.
Nevertheless, the results show that our mechanism is a
viable candidate for explaining the observations. It is also encouraging
that our $T_c$ is an extremely rapidly varying
function of the parameter values. This may explain why
similar effects were not observed in other $2$-$d$ systems.

A magnetic field weakens our superconductivity mechanism, since
it quenches the strong renormalization of interaction parameters that
leads to the pairing. In Ref.\ \onlinecite{triplet} we estimated the upper
critical field to be $H_{c_2} \approx (\hbar c/e)k_F^2\,T_c/T_F$. At the
fluctuation temperature $T^*>T_c$, one would then expect a field of strength
$H^* = (\hbar c/e)k_F^2\,T^*/T_F$ to have a qualitative effect on the
transport properties. At $T^* = 1\,{\rm K}$, this yields $H^* = 9\ {\rm kOe}$.
This is in very good agreement with the observed suppression 
of the conducting phase by a magnetic field.\cite{kpreprint}

\begin{figure}[tbh]
\centerline{\psfig{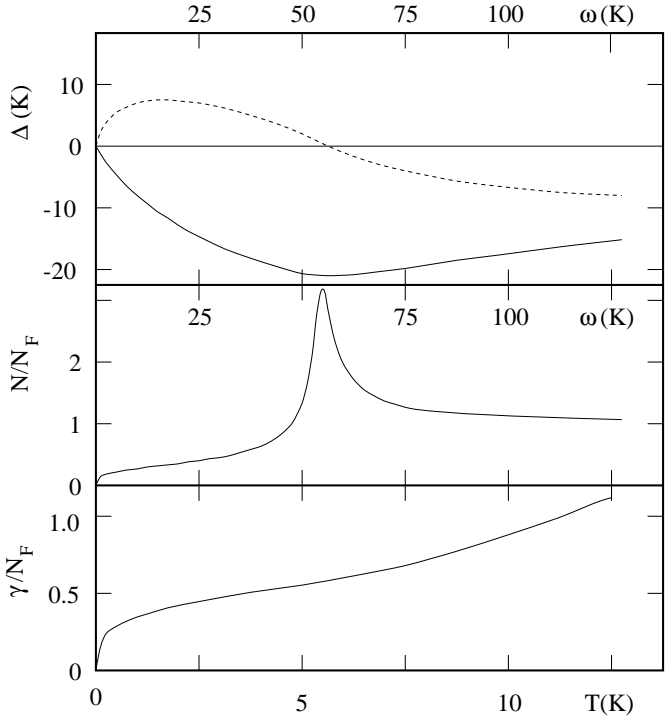}\vspace*{5mm}}
\caption{Real part (solid line) and imaginary part (dashed line) of the gap
 function $\Delta$ at $T=0$ as a function of the frequency $\omega$, the
 tunneling density of states $N$ at $T=0$ as a function of the frequency,
 and the specific heat
 coefficient $\gamma$ as a function of the temperature $T$
 as predicted by the theory for the parameter values discussed in the text.
 $N$ and $\gamma$ are normalized by the free electron density of states per
 spin, $N_F$, and $\Delta$, $\omega$, and $T$ are measured in Kelvin.}
\label{fig:1}
\end{figure}
While the above estimates show that our proposed mechanism is definitely
compatible with the existing experiments, other observations are needed to
convincingly corroborate or refute our proposal. Of particular interest
are the properties of the superconducting phase, if any, that we predict
to exist at sufficiently low temperatures. Apart from the prediction of a
Meissner effect, we have calculated three
observables, viz. the gap function and the tunneling density of
states, both at $T=0$ as a function of the frequency, and the specific
heat coefficient, $\gamma = C_V/T$, as a function of $T$, 
in the superconducting phase. Such calculations
for any purely electronic superconductivity mechanism pose an intrinsic
problem, namely that the feedback of the superconducting state on the
coupling mechanism must be taken into account.\cite{pb,triplet} This
problem has so far not been solved. In the case of the present mechanism
it is likely to just lead to a smaller effective coupling constant
$\lambda = {\hat\Rpersquare}/\pi^2$, i.e. a
smaller ${\hat\Rpersquare}$, then the one that enters the $T_c$ 
equation, except for a qualitative effect in the asymptotically low frequency
regime.\cite{triplet} For our present illustrational purposes we choose
an effective coupling constant $\lambda = 0.4/\pi$, which is a factor of 4 
smaller than what
corresponds to the ${\hat\Rpersquare}$ used for the $T_c$ estimate above. 
The results are shown in Fig.\ \ref{fig:1}. Notice that the purely electronic
pairing mechanism leads to a pronounced pseudo-gap structure on a
frequency scale that is, for our value of the coupling constant, some
5,000 times larger than $T_c$. The $T_c$ scale is reflected in the
region of validity for the asymptotic low-frequency and low-temperature
behavior.\cite{triplet} It turns out that the overall
shapes of the curves do not strongly depend on the effective $\lambda$ chosen,
only the energy scales do. The two relevant energy or frequency scales
are\cite{triplet} 
$\Omega_1^* = (2\times 10^{-2}/\lambda)({\bar\kappa}/k_F)^2 T_F$, 
which sets the overall frequency scale, and
$\Omega_2^* = \Omega_1^* \exp (-1/\sqrt{\lambda})$,
which separates the low and high-frequency regimes 
($\Omega_{1,2}^*$ are the frequency scales
 $\omega_{1,2}^*$ of Ref.\ \onlinecite{triplet} multiplied by $\pi/100$).
In conventional, phonon-induced, superconductivity the role of
$\Omega_1^*$ is played by the Debye frequency, and that of
$\Omega_2^*$ by the value of the gap function at zero frequency.
For the parameter values chosen, the numerical values of the two
frequency scales are $\Omega_1^* = 778\,{\rm K}$, and
$\Omega_2^* = 47\,{\rm K}$.

Figure \ref{fig:1} shows our qualitative
predictions for the superconducting state.
For experiments {\em above} $T_c$, the most promising quantity to measure
would probably be the tunneling density of states, which should reveal
the development of the characteristic structure that is shown in
Fig.\ \ref{fig:1}. Although the full features would be
present only in the superconducting state, one would expect some
structure in the DOS in the region where the transport properties are
strongly anomalous. This structure should become stronger with decreasing
temperature, but not develop into a real gap, in contrast to a BCS
superconductor, but rather into the pseudogap shown in Fig.\ \ref{fig:1}.

We now briefly discuss the spin fluctuation
mechanism, which in principle is capable of producing much higher values
of $T_c$ than the charge fluctuation mechanism. However, for technical 
reasons\cite{triplet} it is harder to
make semi-quantitative predictions in this case.
The $T_c$-formula derived in Ref.\ \onlinecite{triplet} is
\begin{equation}
T_c = T_0\,\exp\left[\frac{-\pi^2}{{\hat\Rpersquare}\gamma_t^0}\,
       \left(1-\frac{1}{32}\,{\hat\Rpersquare}\gamma_t^0\right)\right]\quad,
\label{eq:3.4a}
\end{equation}
Here $\gamma_t^0$ is a bare (i.e., high-temperature) spin-triplet
interaction constant,
and $T_0$ is a microscopic temperature scale that is on the order of the
Fermi temperature. If we assume $\gamma_t^0 = 1$, which corresponds to a
high-temperature spin susceptibility that is twice that of free fermions
with mass $m^*$, and the disorder as above, then
we obtain a value for $T_c$ that is a substantial fraction of $T_F$.
This is certainly an overestimate, for reasons discussed in 
Ref.\ \onlinecite{triplet}. However, the point is that the spin fluctuation
mechanism allows for much higher $T_c$-values than the charge fluctuation
mechanism, provided that there are
strong spin fluctuations in the system, i.e., a large magnetic susceptibility
in the normal metal. If the spin susceptibility of the $2$-$d$ electrons in
MOSFETs could be measured, then it would be very interesting to look for
a correlation between an enhanced spin susceptibility and the occurrence of
the metallic phase.
The qualitative features of the other obervables discussed above
are virtually unchanged for the spin fluctuation mechanism.
Explicit calculations\cite{triplet} of the gap function,
the density of states, and the specific heat, yield results very similar to
those shown in Fig.\ \ref{fig:1}, but with the characteristic features
occurring on a smaller frequency or temperature scale.

We conclude by means of two additional remarks.
First, there has been a recent report about a similar conductor-to-insulator
transition in Si/SiGe heterostructures with a much smaller value for the
disorder than in the MOSFETs discussed above.\cite{pfwComment} However,
the interpretation of these data has been called into question,\cite{kReply}
and more experimental work seems to be needed in order to ascertain if
the phenomenon indeed persists to such low disorder. It is
worthwhile to point out that long-time tails exist even in clean Fermi
liquids, and a mechanism for triplet superconductivity similar to the one
discussed above is conceivable. The effect would be weaker, though, and it
would probably not result in a superconducting ground state even in $d=2$. 

Second, there are other possible interpretations
of the experiments, which are not related to superconductivity. For instance,
it was pointed out in Ref.\ \onlinecite{2-loop} that the RG flow equations 
for disordered interacting electrons in $d=2$ contain a
nonperturbative critical fixed point that corresponds to a MIT. Since
the fixed point is separated from the weak disorder regime by a region
of runaway flow, its physical significance was considered doubtful.
However, Castellani et al.\cite{ccl} have recently speculated that this 
fixed point might describe a MIT in $d=2$. In this context we point out
that one way for the fixed point to attain a physical meaning would be
to make physical sense out of the runaway flow, e.g.
in terms of a non-Fermi liquid state at small disorder.\cite{ac} Indeed, 
in $d=3$, these runaway flows have been related to the 
ferromagnetic state.\cite{fm} The authors of Ref.\ \onlinecite{ccl}
have also argued that even in the absence of a true MIT, the known
flow equations\cite{f,R} for disordered interacting electrons in $d=2$ 
allow for metallic behavior over a sufficiently
large temperature region to explain the experimental observations.
This would require, however, a large spin susceptibility in the metallic
phase, large enough to trigger superconductivity via the spin-triplet
mechanism according to our estimates. Thus 
a combination of some of these suggestions is conceivable: The
pre-asymptotic analysis of Ref.\ \onlinecite{ccl} could explain
the existing experimental data, but at lower temperatures a transition
to the superconducting state discussed in this paper could occur.
In either case, experiments looking for
an enhancement of the magnetic susceptibility would be very interesting.

\acknowledgments
We thank E. Abrahams, S. Chakravarty, V. Dobrosavljevic,
P.~A. Lee, P. Phillips, D. Popovic, and M. Sarachik for helpful discussions.
This work was initiated at the Aspen Center for Physics, and
supported by the NSF under grant numbers DMR-95-10185
and DMR-96-32978.

\end{document}